\documentstyle[12pt,epsf]{article}

\newcommand{\bc}{\begin{center}}

\newcommand{\ec}{\end{center}}

\newcommand{\bd}{\begin{displaymath}}

\newcommand{\ed}{\end{displaymath}}

\newcommand{\be}{\begin{equation}}

\newcommand{\ee}{\end{equation}}

\newcommand{\ba}{\begin{array}}

\newcommand{\ea}{\end{array}}

\newcommand{\bea}{\begin{eqnarray}}

\newcommand{\eea}{\end{eqnarray}}

\newcommand{\bt}{\begin{tabular}}

\newcommand{\et}{\end{tabular}}

\newcommand{\bp}{\begin{picture}}

\newcommand{\ep}{\end{picture}}

\newcommand{\bfi}{\begin{figure}}

\newcommand{\efi}{\end{figure}}

\begin{document}

\title{\huge \bf {Model for Laws of Nature with  Miracles.}}

\author{
H.B.~Nielsen ${}^{1}$ \footnote{\large\, hbech@nbi.dk} \\[5mm]
\itshape{${}^{1}$ The Niels Bohr Institute, Copenhagen, Denmark og}\\ 
\itshape{CERN, CH1211 Geneve 23, Switzerland}}

\date{}

\maketitle

\begin{abstract}
Vi frems{\ae}tter en model for naturlove af M. Ninomiya og mig, i hvilken 
begyndelsesbetingelserne
for Universet er arrangeret if{\o}lge et princip om minimering af en 
st{\o}rrelse kaldet ``imagin{\ae}rdelen af virkningen'' og er defineret som 
en tidstranslationsinvariant funktion (faktisk funktional) af tidsudviklingen 
af universet (alts{\aa} defineret 
p{\aa} systemet af ``veje'' i Feynman-path integral forstand) . Modellen 
synes f{\o}rst 
at give for mange ``mirakler'', men vi kan bortargumentere en masse af dem. 
Dog 
foresl{\aa}s, at SSC-accelleratorens standsning var et eksempel p{\aa} 
et (``anti'')mirakel, der 
virkelig blev set.  
\end{abstract}

\begin{abstract}
{\bf Model for Natural Laws with Miracles}
We review a model for natural laws by M. Ninomiya and myself, in which the 
initial conditions for the Universe are arranged according to a principle 
of minimizing a quantity called the ``Imaginary part of the action'' and 
defined as a timetranslational invariant function (really a functinal)
of the timedevelopment of the Universe (a path way in the Feynman path way
integral sense). The model seems at first to give too many miracles, but 
we can argue away a lot of them. Nevertheless we propose that the interuption 
of the SSC-accellerator were an example of an (``anti'') miracle, that were 
indeed seen. 
\end{abstract}

\section{Introduktion}
The word miracle means approximately an event so strange that it is seems 
in disagreement with the laws of nature. Thus to speak of a system of laws of 
nature which allows some miracles, becomes - it looks - contradictory.
Nevertheless it is precisely an attempt to make a model for how the laws of 
could be such, that they allowed happenings or even led to such happenings, 
which  we would reasonably call  miracles, that is the mainsubject of the 
present article. 
The law of nature that should be changed relative to the usual belief 
to come to the system of laws I want to describe is that nothing 
gets organized in 
the world so as to full fill any simple condition at a LATER TIME.
That is to say that one normally in physics assumes that there is nothing 
that so to speak gets organized with a purpose in mind from the side of the 
laws of nature.
It is therefore one has the second law of thermodynamics  - to which I shall
return - and hich says that when we include the atoms and even smaller 
parts, of which we are built up,then there becomes more and disorder as time
passes. 
According to the usual theory it is namely so, that there must have been 
organized a lot simple properties for the state of the universe in the 
beginning.  
However, there shall not be something, which is arranged or made especially,
so as to guarantee the happening of something specific later. 
When human beings attempt to arrange something so that something planned 
happns they mannage to make some of the simple initial conditions to develop 
so that it becomes just what these human beings are interested in. 
It is the law which I here talk about that says that there is no place for 
a God.
Well, I think I at least met  one case of a colleague, which became 
(or excused 
himself of being) ateist on the basis of this law.

I work mostly together with my colleague Masao Ninomiya\cite{Masao}
\cite{Jorgensen} on a physical model (i.e.on an idea about how we could 
imagine the laws of nature to be) which allows a heading  for special 
happenings in the future - in contradiction with the usually assumed laws of 
nature.       
I have had similar thoughts on non-locallity 
\cite{nonlocalfirst}\cite{nonlocalsec}\cite{vacuumbomb1sv}\cite{nonlocalthirdth}
in connection with what we called 
``the vacuum bomb''
\cite{vacuumbomb1sv}\cite{vacuumbomb2th}\cite{Inf}
(somewhat discussed \cite{Inf2}) 
It really grew out, however, of attempts to justify what we call the 
``multiple point 
principle'' \cite{MPPfirst}
\cite{MPPsecth}\cite{MPPthirdel}
, which roughly states that the vacuum is at
multiple point somewaht analogous to the triple point of water which is the 
combination of pressure and temperature at hich you can have fluid water 
vapor and ice together. We wanted to deliver an argument for that 
the coupling constants and masses (i.e. the parameters of the Standard 
Model or some 
model replacing it) would just adjust themselves to take such values as to 
make several ``phases'' (the different phases of the water in the anlogue 
are the three phases: fluid, vapor, and ice.) able to exist together 
in a kind of balance. Without going too deep in this story of seeking to
argue for having an adjustment analogous to the triple point of water 
let me just say that in our case of hoping to get the coupling 
constants adjusted it looked that an influence from the future were 
needed and helpful.  
Our model is closer to a by a God governed world than the usual theory.
As I shall sketch below we can call our model the 
``model with complex action'', because it consists in that we allow the 
action to have an 
imaginary part (but it requires of course then that I give at least 
a weak idea 
about what the action is).  

The probability that the theory of mine and Masao Ninomiya should be true 
when it in fact is in disagreement with the otherwise generally approved 
principle that nothing truly prearranged - would presumably be so low that 
we would barely have interest to work on it , and presumably hardly 
would have the courrage to talk about it, if it were not actually so that we 
had an example of a ``miracle'' which suits wonderfully into our model\footnote
{I must, however, better admit that we did work on it before we had the idea 
about the closing of the SSC machine being a `miracle´´}:  
One closed  - after having spent huge expences - an accelerator in Texas.
This accelerator should have been able to bring protons to run so fast and 
with so big momenta that they by collission with other ones running the 
opposite way could get sufficient energy so as to according to Einsteins 
equation $E = mc^2$ it could be produced Higgs particles. The Higgs particle 
to which I shall return below, is a new particle after which the high energy 
physicists are eagerly searching but so far have not been able to show exists.
  
One expects to find either the Higgs partcle or something that can replace it.
In our model there are reasons to believe that precisely the production of 
many Higgs particles could be something there would be made miracles to avoid. 
When one had built a quarter of that tunnel in which the particles should rush 
arround with their for single particles huge momenta and energies, well then 
the U:S. Congress closed the project, and now there is only produced a 
bit of champignons in the tunnel!                

Much theoretical physics has the goal of unifying laws of nature, which we 
already know so as to write them in more simple and beuatiful way.  
One may for example claim that Newton with his mechanical laws got united
the Keplerian laws valid in the sky, in heaven, with the earthly laws for 
falling objects on earth to a more general theory, Newtons Mechanic. 
In an analogous way Ninomiya and I could claim to that we are about to unite 
laws for ``equations of motion´´ and ``initial conditions´´ to a united 
theory, which we could call the theory with complex action; but here I must 
then first remind you how one after Newton in physics seperate the informations
 between ``equations of motion'' and ``initial conditions''.  
   

This seperation between initial conditions and equations of motion is in fact
what we shall describe in section 2, as a basis for how we - Ninomiya 
and I - attempt to unite equations of motions with initial conditions.

In section 3 I shall attempt to give reader an idea about what 
the ``action´´ is. 

In section 4 I shall present the main ideas of our model, and at this point 
of the presentation it will look that out model predicts far too many miracles,
so that it cannot be true.  

So it becomes our job at least to presnt some speculations, which could give 
hope that our model after all does not give so many miracles - or that it would
be seen as accidents and not as the government of God, if we shall be able to 
rescue our model from being totally falsified. This is presented in sectio 5. 

Our model means that it will be first of all the quantitativly very big 
happenings - such as the expansion of the universe or about how many neutrons
(the neutron is one of the particles in the atomic nuclei) did survive - that 
are important for how the initial conditions get fixed in our model. We 
shall thus see in section 6,that our model is promissing w.r.t. giving good 
propoties of the cosmological picture towards which one worked oneself forward.

In section 7 we then come to the already announced miracle: The big machine 
in Texas got its funding stopped so that large amounts of Higs particles 
which otherwise would have been produced got avoided. The side meaning of 
miracle that the event that were prearranged is something good is not 
fitting  this example. 
 It were rather one of the greatest disappointments in the 
history of science. So it should perhaps rather be called an anti-miracle 
or negative miracle than just miracle.   

In section 8 we shall then give a resume and say that it is not crazy to 
believe that it could turn out that the natural laws were something in the
direction of our model, so there is an opening for miraculous happenings;
well one could even almost say that we have a model for God - a God than then 
hates the Higgs particles.  

     
\section{Distingtion between initial state conditions \\
and 
equations oof motion}

The law of fall is wellknown: When you release a body (e.g. an aple) in
an in pracsis emty space, it falls in the first secund 9.8/2 meters, by the 
end of the next secund it has reached four times as far down i.e. 4*9.8/2 
meters, and after 3 secunds 9 times so long.  
The rule is: Square the number of secunds and multiply by half the 
(gravitational) accelleration constant, which is the name of the constant
$9.8m/s^2$.  

It were so to speak this law of fall, which Newton united with 
Keplers heavenly laws about how the planets move in the sky.
  
But now we also know that this simple law of fall does not work if one does not
just {\em release} the apple, but instead throws it upwards or slings it 
down. 
In those other cases one can by means of Newtons laws calculate some modified 
laws of fall, but they are different in the detail.
For example the apple thrown upwards might after the first secund be higher 
than the flor from which it were thrown and first later come deeper down.
One shall of course also know when it were thrown or released and from how 
high up, if one wants to express the law of fall by the moment of 
time and the hight over the surface of the earth.  
Such informations about how the development - here the fall - started
are called initial state conditions.
To know how the fall shall go it is neccessary to know both Newtons (second) 
law(s) and the initial conditions. 
Newtons second law allows an infinity of solutions to the differential 
equations, but which of these solutions to choose depends on the initial 
conditions.

Here I want to stress that one after the theory of Newton has divided the 
information about what happens - say how the apple falls or just moves as 
time passes - into two classes of information:

1) The equations of motion, which are differential equations, which must be 
obeyed in order for the motion to be in agreement with the theory (but there 
are infintiely many solutions to the equations of motion - i.e. possible 
functions for the positions of the apple as a functions of time - which obey
the laws of Newton)   

2)Informations about the initial conditions. This is the further information 
that is needed to specify a special one among the solutions, the many 
solutions to the equations of motion. 
Often it will as the name ``initial state conditions'' suggests be the 
velocity and the position of say the apple in a starting moment one has in 
mind as being used to specify these initial conditions. 
But one can also specify a special solution by giving the velocity and 
postion at any moment of time one might want to use.
In fact one can also specify a solution in that way.

We have today a significant amount of information of both these two types.
To {\em unite} several different laws of nature is really what 
especially the group of physicists to which I belong, the 
high energy physicists, seeks to do.   
Sometimes we call it very ambitiously that we search to make a so called 
``theory of everything'' (T.O.E.).
It should mean that one should unite what we know today to a simple and elegant
theory. 
One is in fact close to have such a united theory in what we call 
the Standard Model, but it has several details that simply do not agree with 
experiment. 
In spite of that it seems to be a very good approximation for low energies 
per particle
(one of these small problems for the ``Standard Model'' is that it predicts 
the neutrinoes to be massless, while one in fact has found extremely small
but non-zero neutrino masses. The neutrino is an elementary particle,
that can pass through enourmous amounts of material, and only get stopped 
with very little probability).  
But when the physicists talk about finding the theory of everything 
they usually only think about making a theory for the equations of motion 
modified with quantum mechanics
(which must be used for the particles with which we work;but to make the 
discussion concerning equations of motion versus initial conditions simple,
I believe we shall keep the quantum mechanics outside to begin with, because 
the latter introduces a random behavior of the particles which complicates 
the picture.)  

Here it is then that the ``theory with the complex action'',
as we call it, by Ninomiya and myself will go one step further in 
uniting laws of nature than even 
what our colleagues call ` the theory of everything'': We want also to unite 
the initial state condition with the equations of motion! To make up for it
we postpone to make a theory of everything w.r.t. what the T.O.E.-attempts 
otherwise want: To get the small troubles for the Standard Model cleared up 
by an extended theory.   

About the initial state conditions we have different sorts of knowledge:
When we for instance tell about how the Universe began being very small
and then expanded, then it is information about the intitial conditions.
However it is only information about the intial conditions to tell about 
the situation at one single moment, then one should be able by the euations 
of motion to calculate the situation at e.g. a later moment.

But we have also other information about the initial conditions which we use 
much in pracsis:

Winnie the Pooh ate from a bowl of honney, and would like to eat more to 
test, that there were not cheese in stead of honney in the rest of the bowl.
We often test a little bit of portion of material we have, and then we {\em 
suppose} 
that it will be the same all through. 

But if the Universe were in a quite random state in terms of the small 
particles we are built from, then that of finding some honney would not 
at all guarantee that it should be honney all through.
When in reality one very often is succesfull to find more of the material of
which one find a bit, then it must be because the Unverse is not at all 
in a random state, but in a state in which there are such rules as ``if there
is a little honney, there is likely to be more''.It works well when one 
investigates a bottle, but time reversed it will not work:If you throw a bit 
of honney into a bowl, well, then you cannot rely on that the bowl presumably 
is filled with honney.   

There is in fact for the equations of motion a principle called ``time reversal
invariance'', which states that the equations of motions have such a form that 
a solution for them again becomes a solution if one turns the time (including 
that the particles run the opposite way, as it would look on a film being 
played backwardly.)  
If it is in agreement with the equations of motion that cream runs into the 
coffe and one gets a brown mixture, well, then it must also be in agreement 
with the equations of motion that one starts with the brown mixture and 
at some moment of time cream and coffe seperates again and the cream hops up
into the cream candle, and the coffee left back has become quite black again.
This one however never seems to see realized in nature.So there must be some 
rules for the initial conditions that implies that it cannot occur.

There is in fact a natural law that forbids the brown mixture without 
further intervention just simply to seperate in black coffee and white cream:
This is the second law of themodynamics.

This second law of thermodynamics means that as time passes there becomes 
more and more disorder - provided the particles at the microlevel 
are included-.
If one thus cleans up, so that  there becomes order at some place, then 
ther must be some molecules or other small particles which become more 
disoredered so that the order in total does not fall as time progresses.  
One measures the disorder by the entropy ${\cal S}$ (it is bad luck that 
both the entropy and the action are denoted conventionally by the letter 
${\cal S}$, so that  two expresions for which I wanted to mention 
a mathematical letter symbol in this article happened to be denoted by a 
capital letter S). In fact the second law of thermodynamics says that the 
entropy ${\cal S}$ always grows with time.  

Shortly told the entropy is related to that part of the enrgy which has become
heat and no longer is usefull (for work).
One has in thermodynamics a concept ``the free energy'' $F$ which, if one 
considers 
it already known what entropy is given by the equation
\begin{equation}
F=E -ST,\label{F},
\end{equation}             
but one can of course also use the equation the other way, and get an idea 
about what entropy is, if one first get an idea about what free energy is:
The yet in principle to be made use of energy, the free energy, is what we 
here denote $F$ and equals what one usually calls energy minus the correction
${\cal S}T$, which thus crudely is the part of the energy which is no 
longer usefull to be converted to work.   
When one shall spare on energy for the sake of the environment, it cannot 
be the real energy $E$, that is meant, because that is conserved as a law of 
nature according the the first law of thermodynamics. 
In the equation which I wrote here (\ref{F}) $T$ is the absolute temperature,
which in fact just is the usual temperature in Celcius degrees with $273$
added to the number. 
In a room there is perhaps 300 degrees in absolute temperature, $T=300K$. 
The unit $K$ for Kelvin is to the honor of the Skottish physicist, who 
invented the absolute temperature.
The symbol ${\cal S}$ is as already told a measure for the degree of 
disorder at the microlevel, and it is this disorder that does that we 
cannot anymore get the energy of the molecules say get organized to make 
ordered work.
Second law of themodynamics says that wehen there has first come disorder
into the molecules, so that we cannot govern them to do work, then there is 
no helpfull government of the universe, which will make their motion become
organized, so that we can get other use of it than just heat.
  
This second law of thermodynamics can be said to say that all the order in the 
universe must originate from a very early moment of time, presumably 
the Big Bang-time, and that nothing gets ordered so that there can be order 
in the future which were not already there from the earliest times. Such an 
order which would come in the future but were not already there in the 
beginning would namely be seen as a breaking of the second law of 
thermodynamics: There would in some places be more order than there 
were before. But this would mean the measure ${\cal S}$  the entropy 
(at those places), for the disorder 
would decrease, but that is what the second law of thermodynamics says 
it does not do.
 
 
Ninomiya and I can complain that the usual theory, that it has a law, the 
second law of thermodynamics, that does not obey the principle of timereversal
invariance, and in a way also not the principle of time translational
invariance. 
Can one say that this timetranslation invariance principle is fullfilled
by a natural law as the second law of thermodynamics, which says that all
order has come in in the initial moment?   
It does not sound that the time translation invariance is satisfied,
if the second law of thermodynamics is formulated in this way, that the 
order has come in in an ``initial moment''.  
But this is now a bit of demagogi in order to argue for our model since
as we normally formulate the second law of thermodynamics, ``the entropy 
${\cal S}$ cannot decrease'', then it is in fact time translational invariant.

We must, since the universe has a lot of order in the state in which 
it is,  have some principle leading to this order. 
If we now want to have a very time translational invariant theory, then we 
cannot let this order come in just at one moment of time ( e.g. at Big Bang), 
no, if there once could come order and we have time translational invariance
then it must be possible for order to come in at any moment.
But if there suddenly appear some kind of order that cannot be explained 
in a simple way from the development, then it seems like a government of the 
universe, God one would say. Yes,then  a miracle happens.   

We could for instance imagine - and that is what we do in our model - 
that there one or more types of particles, which this ``governor'' of the
universe would avoid.
With time translational invariance it would have to be so that if this 
``governor'' avoids Higgs particles at one time then he must avoid it at
all times, also to day. 
It would then mean that one should be met by essentially miraculous 
bad luck if one were courageful enough to attempt to produce Higgs particles.
We shall return to such a case of miraculous bad luck in section 7.

\section{The Concept of Action.}
It is not truly important for the understanding of our ideas whether we
formulate our as appearing from a description by means of the action.The latter
is a central function or better functional by means of which one traditionally
describe physical models\cite{Goldstein} :
When the physicists describe their models they do it by means of 
an expression for the (real) action, because it easier to formulate a 
model by writting its (real) action than by writting all its equations 
of motion. 
Then oe can let the colleague write down the equations of motion 
deriving them from the action written down. 
Nevertheless I mean that formulating our model starting from a complex action 
has some motivating value. It is seen when one starts from the concept 
of action and let this action which normally is a real number 
( a real function or better a real ``functional'' of a thought motion) be 
allowed to have 
an imaginary part (i.e.to be complex). 
Hereby I mean that formulating the model this way almost makes it sound as
if we make an assumption less rather than making an assumption more.
This should in principle make it more trustworthy.
In order to give an idea about what sort of mathematical object the 
action is I must ask the reader to think of all thinkable developments 
of the position of an aple moving in space under e.g a throwing of the aple.
One shall not restrict oneself to only think about those those ways the aple 
could be at different postions in differnt moments to those ways which would 
be in agreement with the laws of Newton (the equations of motion), no 
one should even think about those motions which are not in agreement 
with the equations of motion.   
We have an enormous lot of these only thinkable but not  even possible 
motions.
In stead of as Newton did to describe a theory by its equations of motion
the physicists in or days could describe it by providing the expression 
for the ``action''. 
I.e. he would write down the mathematical expression for calculating a 
quantity which we call the ``action'', and we denote by a capital 
${\cal S}$, which depends on a though motion. 
From each of the thinkable motions of the aple it shall from the formula 
which the physicist provided for the action be possible to calculate 
a real number - i.e. a usal decimal number-. From this formula for the action 
the colleague can relativly easily write down the equations of motion for that 
theory to which the action corresponds. 
For the application of the action applies the rule: 
One shall calculate the change of the action due to a very little change 
of the thought upon motion. This becomes then a very small change in the 
action.  
It becomes then a very small change in the action, but the ratio of 
the little change in the action relative to the little change in the 
thought motion does not have to be small. 
Some times, however, even this {\em ratio} becomes zero for certain thought 
motions( used as start for the little change)..  
i.e. that for some very few relative to the huge number of thinkable
motions is the variation of the action by changing the thought about 
motion on which it depends exceptionally small.( Even though it is few 
compared to the toal amount of thinkable motions it is still a huge number
in an absolute sence).
It is just these special thought motions with zero variation of the action
which by definition of the rule of application of the action are those motions
which obey the equations of motion.
It is not difficult to write an action which according to the rule 
for application of the action to vary the action and put the variation to zero
gives the equations of motion of Newton, i.e. gives his laws.
It is also relatively easy to get the equations of motion out of a formula 
for the action, so it is relatively easy to work with the action concept.

It has even turned out that also quantum mechanics has a very nice formulation
called ``the Feynman path way integral'' which exactly makes use of the 
action.
Very crudely we can say that the quantum mechanics means that those particles
we are built from are {\em both partcles and waves at once}.
Light is waves in electric and magnetic fields, but one can also think of 
light  as particles: it moves in straight lines until it hits something - 
and it has turned out that light come in small portions always an integer 
number of them.   
But the quantum mechanics says both theories are true, but that a bit strange
because waves and particles are quite different concepts.
Therefore  even many of the physicists who got Nobel prizes for 
contributions to quantum mechanics did not like it and argued against it. 
For example Einstein argued against quantum mechanics in famous discussions
with Niels Bohr, which contrary to Einstein were one of those who defended it.
Since it according to the quantum mechanics thory seems as if the particles
goes through several slits at once or are several places in the atom at once,
when nobody sees it, it is perhaps not strange that Einstein and others have 
had something to criticize.   

But quantum mechanics have in spite of these attacks turned out to agree
wonderfully with the experiments and we must rely on that it is the true 
theory so far.

With the purpose of argueing for our own model uniting the equations of 
motion with the initial conditions, I shall give an idea of a very beuatifull
formulation of this experimentally very successfull, but philosophically
criticised, quantum mechanics dur to R. P. Feynman \cite{Feynman}: 


The cantral element in this formulation of Feynman is the so called
´Feynman path integral'. 
An integral is allways a limit for a sum, and seen this way the `Feynman 
path integral' is a sum with one term for each of the mentioned 
`thinkable motions'.
We shall thus in principle evaluate a certain 
quantity, the integrand, for every  `thinkable
motion', whether it obeys the equations of motion or not. 
And then we shall roughly speaking add all the integrand results for 
all the `thinkable motions'. 
One could say that that a `thinkable motion' is a path along which the
apple could be thought to be moved, so it is natural that the 
`thinkable motions' are called ``path''(or ``path ways'' ). The integrand 
which is the quantity  which we shall add or integrate over all the 
`thinkable motons'/ `paths', is 
\begin{equation}
\exp(\frac{i}{\hbar} S[path])
\end{equation}                 
    It is important for us here:
1) The integrand is an expression , that depends on the ``action'' 
${\cal S}[path]$,
which again depends on and can be calculated from the `thinkable motion',
here written with the symbol
$path$.
It shall of course be an expression that can be calculated depending on the 
$path$, over which we shall integrate or sum up. 
2) There is an $i$ , which means the formal square root of $-1$, which is
the ingredience in making complex numbers starting from the real numbers. 
If one only allows the real numbers, then $-1$  or negative numbers alltogether
have no square roots.
A square root $i = \sqrt{-1}$ is not truly existing , but ``imaginary'',
fantasy, or formalism. 
When one allows such an $i$ one works with the complex numbers instead 
of just the real ones.
Due to the appearance of this $i$ the integrand which we should 
integrate or add up over all the `thinkable motions' or paths thus 
a complex number( and not real).
3) The appearnce of $\hbar$ is a sign that it is a quantum mechanical 
toery, because $\hbar$, the so called Planck constant, is a natural 
constant which annouces that it is quantum mechanics we use. 
In high energy physics we are usually using units so that we put 
to unity, so that we in pracsis ignore it, untill we have completely 
finished and shall translate the results to the usual units.
In fact $\hbar$ is an enourmously small number in usual units, and when it 
comes in the denominator like here, we get an enourmously large 
number, thus  
$\frac{i}{{\hbar }}* S[
path
]$ 
is typically an enourmous number.
             
 \section{Our Model with Complex Action} 
\subsection{Our argument by using `Feynman path integral'}
There something relatively strange in the the usual `Feynman path way'
formulation of quantum mechanics:
The presumably most important quantity in the `Feynman path way integral',
the integrand, which is what we shall sum or integrate, is a complex number,
but the quantity from which it is computed is the action ${\cal S}[path]$
which real.
One could ask:
If something should be limited to only be allowed to be real and not be 
allowed to be complex, should it then not be the most important quantity 
itself,the integrand, and not a quantity we find deep inside the expression 
for this integrand? 

To take the integrand itself to be real would go quite wrong by using the 
`Feynman path way integral', but what if both the action and the integrand
were allowed to be complex?
It looks at first also not so promissing, but this is actually our model.

\subsection{About How we work with our Model}
Since the usual quantum mechanics and the usual classical mechanics 
 which can be derived form the usual `Feynman path way integral'
function very well with the real action - this is the usual theory -
it is our task , if we want to defend that the action fundamentally 
is complex, to explain away all effects of the action being complex if it were.

It looks promissing that we can explain away the most obvious revelations 
of the complex action.
We attempt to show that one gets in an approximation classical equations
of motion as usually obtained using just the real part out of our model,
without the imaginary part having any significance for the equations 
of motion.
I.e. we hope that we do not spoil the equations of motion by allowing 
the complex action in that approximation in which we ignore quantum mechanics.
This is already a great victory for our model.

What turns out to be the main effect of the imaginary part of the action,
which we have allowed, is that it chooses which solutions shall most 
likely be realized.
It comes from that the imaginary part of the action comes to deliver 
a probability wieghting, which selects some motions - in fact among the 
solutions to the equations of motions - to be much more probable than 
others!  
Thereby the imaginary part of the action come to play the role of determining 
that part of the informations which we above classified as the ``inital 
state conditions''.
Our model thus develops into a model that gives both the initial state 
conditions and the equations of motion.
It can be said to be a unification of these two domains of informations,
provided of course that it would function as a theory.

It turns out to be the rule determining which classical solution to the 
equations of motion shall be realized in our model  - that it shall be that one
that has the smallest (in the sense of most negative) imaginary part of the 
action.
We remenber that the characteristic postulate of our model were
that the action ${\cal S}[path]$ is a complex number, which thus have two
parts, the usual real part and the imaginary part, which comes with the
factor $i = \sqrt{-1}$ in front of it. 
It is this latter part, the imaginary part, which is the new thing and 
which gets significance by determining which one among the infinitely many
solutions shall get realized. 
One could then imagine that one found all solutions to the equations of motion
and for each of them calculated the imaginary part of the action - we think it 
is given by an expression we think we have been able to guess with support 
in the experiments, like one knows  the form of the real part of action 
for the Standard Model. The realised motion should then in our model be 
that one which gave the smallest imaginary part of the action.
One sees really here that our model is very ambitious:
If one could - what one of course normally cannot - calculate so as to find 
the solution with the minimal imaginary part of the action, then one would 
not only have the usual mechanical equations of motion, which gave 
restrictions how the motion could occur so that it obeys Newtons second law,
but in fact predict it all also how the apple throwing starts. 
This means that in principle that a theory of thetype of our model 
is a ``theory of everything'' to a greater extend than those ones my 
colleagues hope for, because we also should be able to answer what should 
happen to all times. 

Practically such calculations are of course totally impossible, except 
that one might be able to get something out in by some clever method 
in exceptional cases. 

\section{Do we get too many Miracles?}

Let me attempt to resume which type of model we have made:

It is a model that givesa mechanics with the usual mechanics or dynamics
with the usual equations of motion.
But then there is further that rule that which solution shall be realised 
is determined as the one giveng the smallest number for a certain quantity,
i.e ``the imaigary part of the action''. 
The latter can be computed according to  a relatively simple formula.
The central point is not if this quantity is especially beuatifull because
it is the imaginary part of the action, but the fact that one in principle
in the model have a calculational method - namely to minimize ``the imaginary
part of the action''   - from which one can determine the initial conditions.
It corresponds approximately to that the World is governed by a director,
who seeks to minmize ``the imaginary part of the action'' in an analogous way 
to a leader in the industry seeking to minimize the deficit of the firm.
He would seek to avoid strongly loss giving businesses, and in the same way 
the Universe in our moedel should avoid e.g. those types of particles,which 
would give big positive contributions to the ``imaginary part of the action''. 
If we e.g. assume that the mathematical expression according to which one 
calculates the ``imaginary part of the action'' that it gets big positive 
contributions from Higgsparticles, well then the Higgs particle will be 
avoided for that solution which gets realised. 

But now we see that with such a government of the Universe we easily come to
see a lot of regularities - one allways have bad luck with the activity,
if one attempts to produce certain type of particles, while one  
to make up for it get surprisingly good luch when producing other 
particles - that it does
not agree with experiment.  

The truth is that the very most that happens in the universe runs so 
that easily could be accidental consequences of how the universe started in
a relatively special state (at the Big Bang time).
We may well have read about a few miracles in the Bible and legends of 
saints, but immediately it seems that our model it looks that our model 
with a formula that gives enourmous probability ratios between solutions
which only deviates by few e.g.unwanted particles, seems to give many more 
miracles of positive and negative character than we have heard about.  

This means that if we shall rescue our model from immediate falsification,
we must discover that we can calculate there will be much fewer miracles 
- or shall we say  non-accidental event - than we would estimate at first.  

\subsection{Suppression of Miracles in our own Era}

There are several arguements which to some extend could explain away
the government effects, so that simple happenings might be organized 
to day (by the laws of nature in our model), such as to prevent the
production of a paricle type giving a positive contribution to the 
``imaginary part of the action'', in occuring in large amounts:

Relative to what the situation were in the Big-Bang-time, when the 
universe were a split second old, the density of matter is to day 
enormously little. 
So to day the universe is empty - and that eveen after human scale:
Human beings cannot pump a container out to  so good a vacuum that 
it can compete even with the average density in the universe.
From the point of view of the director for the development of the Universe,
which is the essence of our model, the present time is less interesting 
than the Boig-Bang-time when there were much particles present per volume.
To day it must be mainly the contribution to the imaginary part of the 
action from the empty space vacuum that counts.

It also contributes to make the time long after Big-Bang less interesting 
w.r.t governing what shall happen according to our model, that those partcles
of which we mainly consist, electrons, protons, neutrons bound in the nuclei,
are in pracsis conservedso that there are equally many of them all the time.


If the ``director'' wanted to avoid them or to have as many as possible, he 
should in fact just arrnge it in thestart, and then there would not be 
anything
miraculous in there being the same number the rest of the time.
 It is a natural law, counted under the conservation laws, that these particles
are conserved in number ( when the neutronns do not run freely arround
but only are present inside the stable nuclei). 

The fact that these of which we consists move with rather low velocity 
relative to the speed of light means that they move very little seen form 
the point of view of the theory, which we shal use to give us both the 
real but also the imaginary parts of the action.
We namely count on that it is the speed of light which is the fundamental
unit for velocity and that it is thus velocities relative to the speed 
of light, that comes into both the real and the imaginary actions. From the 
point of view of the government of the world in our model the particles to day
lie practically stil (relative to the fundametal light speed).
This further contributes to that the present time is dull for the ``director''
and thus it is not worthwhile to spend miracles on it.

One must take into consideration that if something shall be arranged into
a little miracle to day, and the equations of motion shall be fulfilled 
at all times, then it imposesrestrictions on what can be arranged at 
other times, because it restricts the number of solution that can be used. 
The majority of solutions would of course not have the little for to day
organized miracle.
There is simply too much competion between various eras in which the 
``director'' - i.e. the choice of solution - can make miracles without 
the one miracle disturbinmg the other one due to the equations of motions.
It becomes difficult to get a lot of miracles in our own time
which makes up only a very little fraction of even the age of the universe
being 13.6 millard years. 
The time near Big-Bang is more important than our time so most arrangements 
will be for that time.

The argument that the particles have almost stopped in our time 
, from the point of view of the velocity of light, is no more valid for 
those accelerators which the high energy physicists use. 
One might define high enrgy physics by the investigated partices move 
with velocities of the same order as that of light.
This means that other things equal we expect that our will set in with 
governing, more miracles, when we are concerned with high energy physics 
than with dayly life in which the particles normally are slower. 
The early time of the universe shortly after Big Bang were a time with many
particles with velocities of same order of magnitude as that of light.  
It were a high energy physics time.
Some times one talks about the Big-Bang-time as the pure mans 
high energy accelerator.``The pure man'' who lacks the milliards 
to build high energy accelerators as LHC and SSC etc. can study high energy 
physics by studying Big-Bang-time (but nowadays many of the astronomical studies 
of Big-Bang-times is something being performed from satelites, so it is also
not so cheap.). 
 
\subsection{Timereversal invarians in our Model} 

Our model is at least both timetranslational invariant and 
approximately time reversal invariant in the same way as the Standard Model.
For our discussion we can say it is time reversal invariant and we 
could choose it so if we liked.
So one must with worry ask, how it can mannage to be in agreement with 
the second law of thermodynamics?
The answer is that it is the fact that we live {\em after} the Big Bang time
which were the for the selection of the solution so important/interesting  era
that means that the past is ( for us) more organized than the future, which 
does not get organized or arranged, beause it is rather dull from the 
point of view of the ``director''. 

\section{Cosmologically our Model is Promissing}  
The fact that we have the longest time of the universe history 
spent as a rather dull time as described above - almost no particles 
that even move very little - is presumably a good idea for 
obtaining the minimal imaginary part of the action.
For if the system/the universe varied much with time, it would come through
many different states and it would soon get a bit more and soon a bit 
less contribution to this imaginary part of the action.
But it is presumably easier to find just one configuration which gives agood 
negative contribution and then let the system stay there arround.
It is what is achieved by the Hubble-expansion having cooled down the 
universe so strongly that there is alomost nothing that moves compared 
to what happened in the Big-Bang-time. Hubble expansion has attenuated the 
Universe so much that it is responsible for it being so close to be empty. 

We can imagine that it in spite of the above would pay better to have a more 
active time abot Big-Bang, because it could be that there were a configuration 
that gave enormous negative contributions to the imaginary action But most likely
the configuration(s) with the enormous negative contributions would not 
remain as times goes on, and it would benecessary to let the system end up 
to spent the longest times on some reasonably stable configuration like the 
present vacuum. 
The Hubble expansion can be used both having a time with a lot of particles 
which we may think gives an enourmous negative contribution to the 
the imaginary part of the action and a more stable and controlled time 
which should make it possible to get a rather favourable contribution from
almost just one state, the vacuum.
It seems to be a very smart idea from the side of the ``director'' to 
make a universe with a Hubble expansion so as to get both some highly negative 
contribution from an unstable but very strongly contributing era and 
a longer time more stable contribution froma not quite so strongly 
contributing vacuum. 
Our model predicts a Hubble expansion.
This is what we mean by saying that our model is promissing for 
the cosmology. 
 
\section{The Higgs Particle and the Miracle}

As already told the high energy accelerators  - in which one has particles 
with high speed - is more interesting for governing ( in our model) the 
development of the world
than non-high-energy- phenomena. 
It should thus mean that if there is high energy physics on a large scale,
then that should be where we should look for miracles in our model..

\subsection{The Higgs Particle}
\subsection{Higgspartiklen}

Before I tell about the accelerator SSC ( Super Conducting Suoer Collider)
which essentially ``miraculously'' was stopped, I would like to talk a bit
about the Higgs particle, which is expected to be found in the accelerator
LHC(Large Hadron Collider) - it is alomst finished here in CERN in Geneva),
and also in the older accelerator, if the latter come to work.  
 
That theory, the Standard Model, to which one has reached, and which seems to 
agree very wel with experiment for such energies as one has we have reached to day
that many colleagues are almost a bit sorry that it is so good that there is
no use for an improvement ( except for some small deviations from agreeing with
experiment), has a particle in it and which is not yet found, and it is called 
the Higgs particle.      
It has a crucial role to play by causing that many of the other get masses 
different from zero.
Either it has to be there or there has to be some replacement for it.

 
\subsection{The Hierarchy Problem}
In connection with the Higgs particle there is technical problem,
the hierarchy problem, which is connected with the problem why the 
Higgs particle has such a small mass as it needs to have in order to do its job
of giving masses (but sufficiently small also) to the other particles. 
One would without a good explanation have expected a much bigger mass 
for the Higgs particle than what we already know that it must have.
This mass appears in the expression for the action - the mass which we know 
crudely about, and whic is surprisingly small, appears of course in the real 
part of the action.

W.r.t. a corresponding mass ( to the appearance of the square of it in the 
real part of the action)in the imaginary part of the action, appearing 
Higgs-mass parameter(an imaginary part of the square of the mass)
there are now two possibilities: 

a) The mysterious mechanism (making the real part of the  Higgs mass square 
small) can be applied to the imaginary part of the action too, so that its 
corresponding term also becomes surprisingly small.

The mysterious mechnism only functions for the real part of the action,
so that the imaginary part gets a to the Higgs mass corresponding parameter,
which have the {\em a priori expected size}, which is enormously much
bigger.

In this latter case the Higgs mass term in the imaginary part of the action
can be enormous, and we would expect an enormously big interest from the 
side of the ``world director''w.r.t.to exactly Higgs particles. 
Were He really enormously interested in a positive way, i.e. liked the 
Higgs particle and sought to make many of them, then He could have filled 
the Universe with many more of them.
At least it our guess, that He rather hates Higgsparticles, and attempt to 
avoid them.   

But that means, that a solution to the equations of motion - for which 
there at a certain time is being built a big accelerator producing a lot of
Higgs particles - should be very little chanse for being just that soultion,
giving the minimal imaginary part of the action. 
It should thus mean that we do not expect that this type of machines 
will be built in the real world( i.e. on the soluiton which gets realized.).

\subsection{The Miracle}

Now it is not so easy to know, if there is going on some prearrangement,
that ensures, that there is being built much fewer of a type of accelerator,
than there would bebuilt without such an effect, because what would happen 
if the effect were not there in the case it actually is there.   
We have to remeber that at least approximately we have the equations of 
motion fullfilled ( at least classically, i.e. without quantum mechanics),
so that for everything that happens there will be an explanation from using 
the equations of motion.  
The miracles of our model are only prearraged events: 
The initial state conditions have from the earliest time/ Big Bang or even 
before, if there were any before that, been arranged so, that the seemingly 
miracle just come to appear.    
It shall not be miraculous by breaking the equations of motion, but only 
by being a strange coincidence of the development of the particles 
or the events taking place, that the miracle comming up just should
happen.
It shall thus by seeing that what happens does not seem to be probable 
that you shall see that it were a miracle( in our model).  

\subsection{The SSC-accelerator}
In Texas near Dallasthere were an accelerator being built, SSC(Superconducting
SuperCollider) relative to which the LHC accelerator ( Large Hadron Collider)
in CERN in Geneva, to which we look forward with great expectation that it 
shall start this year, would be a dwarf.  
But then in October 1993 the Congress decided to stop the garnds for SSC.
Jefrey Mervis and Charles Selfe have written an article 
\cite{SSCstop} about, why this brilliant accelerator were stopped at a moment
when 30 km tunnel were already built. 
It is not normal
\footnote{There were an accelerator ISABELLE in Brookhaven
which would not havebeen able to produce Higgs particles which were stopped
in 1983 - after having got 23 million dollars in 1979 - after technical 
problems and in favour of SSC ( which as we tell were stopped ten
years later); there is though now under the use of tunnel etc. 
from ISABELLE built an accelerator RICH, which functions fine.}
that one stops such a construction when a quarter of the tunnel is
built.
Even though the price for the DR-city (The new place for the Danish 
State Radio)on Amager also came too high up relative to the plan, one did not 
give up totally to finish it as it were the case of the SSC-accelerator. 
It were a remarkable stop even though therewere many reasons for that it 
failed.

Is it not just such a series of small cases of bad luck, that seperately 
easily could occur accidentally, which would be the easiest way to stop
the machine, without it requiring the truly great miraacle to do it, a sort
of economizing with the ``miraculosity''( if we can use such a word 
fr the degree unlikeliness of the events).   
This SSC would have produced Higgs particles, if it had come to work,
so if there were something to avoid it were the right machine to stop. 

\subsection{Miracle - not quite the right word}
The just described ``miracle'', the stopping of the SSC-accelerator, were
at least not a miracle , if one to `miracle' associates the further 
significance, that a miracle shall be something {\em good}. 
This stopping of SSC were rather a catastrophy for science and the 
economy of the surrounding region.
Vi likely need a new word, a ``negative miracle'' to denote this 
type of events, which are so strange, that they ought strictly speaking 
to be classified as 
miracles, but absolutely not are miracles in the sense that they 
are events, which we want to happen. 

If our model could be shown right by this ``negative miracle'', it could 
perhaps turn out to give science more than the particles for which we 
missed to find by means of SSC.
It could give a kind of theory about ``God''. Then it would perhaps
both a strange {\em and  good} event, that could deserve to be called 
a miracle. Also with the side significance that it were good.  

\subsection{LHC and Tevatron ?}

Now soon the LHC shall also produce Higgs particles, and perhaps 
the Tevatron, an accelerator in Chicago, for the time the biggest 
of this kind of acclerators, it already though without we really knowing 
if it does.
It is in fact a bit worrisome for the truth our model/theory that this 
Tevatron presumably already has produced thousands of Higgs particles, 
though without having been able to get a single one statistically convincingly
detected.  
May be we can only rescue our model by saying that it is not the Higgs particle
but a quite differnet process, which SSC would have caused, that were the 
reason for that it called for its own bad luch in our model.
The breaking of the baryon number conservation is one of the effects in SSC 
which were
discussed theoretically.
It means that the number of baryons(a class of particles to
which the proton and theneutron belong) minus the number of their 
anti particles,antibaryons, possibly opposite to the case usually 
when it is constant, would change a little bit in SSC.

Masao and I have written an article 
in which we propose that we 
should play a game about the destiny of the LHC (the accelerator in CERN).  
so that if a special card combination is pulled by the Director General
for CERN, then the LHC should be restricted concerning how many Higgs 
particles it would be allowed to produce.
The idea then should be that if our model were really true then the 
realized solution to the equation of motion would presumably turn 
out to be one wherein the seldom combination giving restictions were pulled,
rather than one in which a lot of small accidents should make the machine 
closed.

If such a seldom combination were pulled it would be a clear 
test of our model.
It is much more difficult to calculate how unlikely were the 
combinations of bad luck, that hit SSC, than to compute the 
probability for a combination of cards in a card-pulling. 

\section{Conclusion and Resume}

I have put forward a model (by M. Ninomiya and myself) which 
in {\em one} model - the model with ``complex action'' -
include both equations of motion  and initial conditions - or raather,
and that is the new in it: It determines by an in principle possible 
calculation, which among the infinitely many solutions to the classical 
equations 
of motions in fact gets realized.  
In principle, but not in pracsis, one could ask our model approximately 
about everything, because it truly talks about everything! The other so called 
``theories of everything'', which the highe enrgy physicists hope for, are 
not usually so ambitious, that they even want to predict the initail 
state conditions.   
In fact the principiel way of calculating how the Universe starts is the
following:

For all solutions to the equations of motion - one can construct one 
each time one delivers a set of velocities and positons for each degree 
of freedom - one calculates the imaginary part of the action.
That solution which gives the smallest (the most negative) imaginary part
of the action is the right one, i.e. the one that is realized.

The important thing is not so much which function (strictly speaking it is 
called a functional when it is function of function, such as a function 
of a path), which one shall minimize, but that one has such a calcualtional
method in principle. It is nemely that which means that the model 
predicts everything.

In pracsis it can be more difficult to apply the model - and one shall at least
have guessed or fitted the parmameters in the expression for the imaginary
part of the action in order to be able to calculate on it - so one must 
make statistical estimates instead.
But if now there were a very big effect, as we see the possibility for,
when we are concerned with the Higgs particle, then it could be that there were
so big effects, that one could see miraculous effects without quite eneormous 
and impossible calculational work.  

I have in this way argued for that it were very likely an effect 
of our model, that the big accelerator in SSC which were partly built 
in Texas nevertheless in 1993 were stopped by the Congress of the 
United States. 
I consider this event as a sort of (negative) miracle.
The ``God'' or ``director'' which in reality is in our model and which adjusts 
the initial state conditions for the universe ( the start) such that the 
imaginary part of the action becomes smallest possible, would like to avoid 
the Higgs partiles which the SSC-machine would have produced if it had 
come to work.
Thus He let a series of smal accidents hit the machine, so that it got 
stopped, before it produced any Higgs particles. 
Such a break of the development of the machine is so remarkable,
that we can denote it a negative miracle.
for det var en katastrofe for videnskaben og omr{\aa}det, en af de st{\o}rste 
fiascoer af denne art.).

In the time of writting we wait upon whether the little brother of SSC,
the LHC will be alowed to staart, or whether it will also be so hated
by the director which effective is present in our model that it will
also be stopped.
\subsection{Note added in last moment of translation}
When this translation and its submission is being finished, there has already
been held a formal inaugoration celebration for LHC, {\em but} there were 
leak on a helium  tube so that the start of true functioning of the 
machine would at least be delayed by two month. To day, however, I heard that 
indeed the failure were a bit more severe and the machine would first come to
function again after the winter closure - one has to close in winter because 
the electricity is more expensive in the winter -. Indeed it is now expected to
start in May 2009. 

It should also be brought in mind as possible mechanism for closing the 
machine that there has been some attempts to make the LHC declared 
dnagerous because of the possibility of producing small black holes
or making strangelets; such objects should be able to an erradication of earth.
There are however in the Lsag report convincing argments from that 
if so the moon and earth would already have suffered such a fait just
from the cosmic radiation. Even though there is thus no real danger at all,
the mere fear against such danger could potentially cause a thread to 
the running of the machine. Also there were a sue against it because of 
these speculative - but unexisting - dangers.

\end{document}